\documentclass[12pt]{iopart}
\usepackage{iopams}  
\usepackage{graphicx}

\newcommand{\myvec}[1]{\vec{#1}}
\newcommand{\vk}{{\myvec{k}}}
\newcommand{\vr}{{\myvec{r}}}
\newcommand{\vA}{{\myvec{A}}}
\newcommand{\vnabla}{{\myvec{\nabla}}}
\newcommand{\Ef}{\mathcal{E}}
\newcommand{\Rth}{{\theta R_0}}
\newcommand{\vEf}{\vec{\Ef}}

\newcommand{\ddt}{\frac{d}{dt}}

\renewcommand{\r}{\rangle}
\renewcommand{\l}{\langle}
\newcommand{\up}[1]{ ^{(#1)}}

\newcommand{\beq}{\begin{equation}}
\newcommand{\eeq}{\end{equation}}
\newcommand{\bea}{\begin{eqnarray}}
\newcommand{\eea}{\end{eqnarray}}
\begin{document}

\title[t-SURFF: photo-electron spectra from minimal volumes]
{Photo-electron momentum spectra from minimal volumes: the time-dependent surface flux method}

\author{Liang Tao$^{1,2}$ and Armin Scrinzi$^1$}

\address{$^1$Ludwig Maximilians Universit\"at, Theresienstrasse 37, 80333 Munich, Germany}
\address{$^2$Wolgang Pauli Institute c/o University of Vienna, Nordbergstrasse 15, 1090 Vienna, Austria}
\ead{armin.scrinzi@lmu.de}
\begin{abstract}
The time-dependent surface flux (t-SURFF) method is introduced for
computing of strong-field infrared photo-ionization spectra of atoms 
by numerically solving the time-dependent Schr\"odinger
equation on minimal simulation volumes. The volumes only need to accommodate the electron quiver 
motion and the relevant range of the atomic binding potential. 
Spectra are computed from the electron flux through a surface, beyond which the 
outgoing flux is absorbed by infinite range exterior complex scaling (irECS). 
Highly accurate infrared photo-electron spectra are calculated
in single active electron approximation and compared to literature results. 
Detailed numerical evidence for performance and accuracy 
is given. Extensions to multi-electron systems and double ionization are discussed.
\end{abstract}

\maketitle

\section{Introduction}

In a broad range of recent experiments, strong infrared laser pulses, often combined with high harmonic pulses, 
are used to study the electronic dynamics of atoms and molecules on the natural time scale of valence electron motion 
of $\lesssim 1\,fs$. Basic mechanisms of the IR-electron interaction are well understood within the simple
semi-classical re-collision model \cite{corkum93:simple-man}, but for a more detailed understanding numerical simulations must be employed.
This is due to the fundamentally non-perturbative interaction of near IR fields with valence electrons at intensities
of $\gtrsim 10^{14}W/cm^2$. Even for the simplest single-electron models the simulation remains challenging,
especially when accurate photo-electron momentum spectra are required, as, e.g., for re-collision imaging 
\cite{spanner04:diffraction,yurchenko04:diffraction,meckel08:diffraction}.
When two-electron processes are involved, one quickly reaches the limits of present day computer resources 
\cite{taylor03:780nm,martin07:science}.

The surprising difficulty in simulating a seemingly simple process like ionization by a dipole field is due to 
the presence of vastly different length- and time-scales: first, even though the laser pulses
can be as short as a single optical cycle, at the Ti:Sapphire wave length of $\lambda=800\,nm$ this still corresponds to 
a FWHM duration of $\gtrsim 2.5\,fs$ or about 110 atomic units (a.u., $\hbar=e^2=m_e=1$). 
During that time, a photo-electron with an energy of $13\,eV \approx 1/2\,a.u.$  moves to a distance of 
$\approx 110$ Bohr, which sets a lower limit for the required box-size, if reflections from box boundaries
are to be avoided. In practice, higher energies and longer pulse durations including the rise and fall of the 
pulses are of interest, leading to simulation volumes with diameters of thousands of  
atomic units. At the same time, photo-electron spectra are broad, extending to at least $2\,U_p$ for the 
``direct'' photo-electrons and further up to about $10\,U_p$ for the ``re-scattered'' electrons. Re-scattered
electrons are those that after ionization re-directed to the nucleus by the laser field, where they absorb more photons in 
an inelastic scattering process. The ponderomotive potential $U_p=I/(4\omega^2)$ grows linearly with 
laser intensity $I$ and quadratically with wavelength. At the moderate intensity $I=10^{14}W/cm^2$ 
and $\lambda=800\,nm$ it is $U_p=0.22\,a.u.\approx6\,eV$. The re-scattering momentum energy cutoff of $10\,U_p$
corresponds to a photo-electron momentum of $2.2\,a.u.$ For representation of such 
momenta on a spatial grid, we need grid spacings of at least $\Delta x \lesssim 2\pi/2\,a.u.$, for accurate
results usually significantly more than this. This leaves us with thousands of grid points in each 
spatial direction even for moderate laser parameters. The situation quickly worsens at 
higher intensities and longer wavelengths.

This general requirement on discretization cannot be overcome by any specific representation of the wave function: 
speaking in terms of classical mechanics, we must represent the phase space that is covered by the electrons, 
which involves a certain range of momenta and positions. If we have no additional knowledge of the structure of 
solution, the number of discretization points we need is the phase space volume divided
by the Planck constant $h$. In some cases like, for example, single-photon ionization, we can exploit the fact that
at long distances the solution covers only a very narrow range of momenta and only the spatially well-localized
initial bound state requires a broader range of momenta: the phase-space volume remains small, simple models
like perturbation theory allow reproducing the physics. We have no such simplifying physical insight
for strong-field IR photo-ionization.

The lower limit for the number of discretization points for the complete wave function can be 
approached by different strategies: the choice of velocity gauge \cite{cormier96:gauge}, working
in the Kramers-Henneberger frame \cite{telnov09:kramers-henneberger} or in momentum space \cite{zhou11:p-space}, by variable 
grid spacings, or by expanding into time-dependent basis functions \cite{scrinzi01}.
A promising strategy is to follow the solution in time \cite{hamido11:time-scaling}.

Alternatively, we can abandon the attempt of representing the complete wave-function and instead use 
absorbing boundaries and extract momenta at finite distances. The time-dependent surface flux method
(t-SURFF) introduced here is such an approach. After its mathematical derivation, numerical implementation is briefly 
discussed. Angle-resolved photo-electron spectra are presented using between 75 and
200 radial discretization points for truncated and full Coulomb potentials. We discuss
accuracies and demonstrate the efficiency of t-SURFF by comparison with recent literature. Finally,
possible extensions to few-electron systems and double photo-electron spectra are outlined. 

\section{The t-SURFF method}
Scattering measurements and theory are both based on the plausible 
idea that interactions are limited to finite ranges in space and time and that 
at large times $>T$ and large distances $>R$ 
the time-evolution of the scattering particle is that of free motion:
\begin{equation}\label{eq:asymptotic-expansion}
\Psi(\vr,t)\sim \int dk\up3 \exp(-it\vk^2/2) b(\vk) \chi_\vk(\vr)\quad{\rm for}\,t>T,|\vr|>R,
\end{equation}
where $\chi_\vk(\vr)=(2\pi)^{-3/2}\exp(i\vk\cdot\vr)$ are $\delta$-normalized plane waves.
The measured momentum spectrum is proportional to the square of the spectral amplitudes $b(\vk)$
\begin{equation}
\sigma(\vk)\propto |b^2(\vk)|.
\end{equation}
For Hamiltonians that are time-independent beyond a certain time $T$, we can readily obtain the 
spectral amplitudes $b(\vk)$ by decomposing the wave function $\Psi(\vr,T)$ into its spectral components
\begin{equation}
b(\vk) = \l \psi_\vk|\Psi(T)\r \exp(-iT\vk^2/2),
\end{equation}
where the scattering solutions 
\beq
H(T)|\psi_\vk\r=\frac{\vk^2}{2}| \psi_\vk\r 
\eeq
have the asymptotic behavior 
\beq\label{eq:asymptotic}
\psi_\vk(\vr)\sim\chi_\vk(\vr)\quad{\rm for }\quad|\vr|\to\infty.
\eeq
For computing $b(\vk)$, one needs to (i) propagate a solution $\Psi(\vr,t)$ until time $T$, 
(ii) obtain the scattering solutions $\psi_{\vk}$. Unfortunately, both of these tasks are non-trivial
in all but the simplest cases. As discussed above, solving the TDSE with IR fields 
is numerically challenging because of the large box sizes needed. ``Obtaining the scattering
solution'' amounts to outright solving a time-independent scattering problem for $H(T)$,
but analytic scattering wave functions are only known for simple model potentials 
and for the Coulomb potential. 

Rather than letting the system evolve and analyzing it at the end of the evolution,
we can record the particle flux leaving a finite volume as the systems evolves. Such
a procedure neither requires stationary 
scattering solutions nor the complete wave function in an asymptotically large volume. 
It is practical, if we either know the further time-evolution of the system
outside the finite volume in analytic form or if it can be obtained 
with little numerical effort.
This type of methods has been applied for reactive scattering with time-independent 
Hamiltonians \cite{balint-kurti91:spectra,tannor93:spectra}, where obtaining scattering 
solutions would be tantamount to solving the complete 
stationary scattering problem, a daunting task for few-body systems. 

Photo-ionization cannot be computed by existing surface flux methods, as the dipole 
interaction is non-local and the external field modifies the 
particle energies everywhere, in particular also after the particle has left the 
finite simulation volume. To handle this, we have developed the time-dependent 
surface flux (t-SURFF) method, that will be derived in the following.
A preliminary version of the method was published in \cite{caillat04:mctdhf}.

Let us choose a surface radius $R_c$ large enough that the particle motion
can be considered a free motion and that all occupied bound states of the system
have negligible particle density at $|\vr|>R_c$.
Let us further pick a sufficiently large time $T$ such that all particles that will ever
reach our detector with energy $\vk^2/2>0$ are outside the finite volume $|\vr|<R_c$.
At that time, the wave function has split into bound and asymptotic parts 
\beq
\Psi(\vr,T) = \Psi_{\rm b}(\vr,T) + \Psi_{\rm s}(\vr,T)
\eeq
with
\bea
\Psi_{\rm b}(\vr,T)\approx 0 &\quad{\rm for}\ |\vr|\ge R_c\\
\label{eq:psi-s}
\Psi_{\rm s}(\vr,T):=\int dk\up3 \exp(-iT\vk^2/2) b(\vk) \psi_\vk(\vr)\approx 0&\quad{\rm for}\ |\vr|\le R_c.
\eea
The approximate sign in (\ref{eq:psi-s}) refers to the fact that very low energy particles $\vk^2/2\sim 0$ 
may not have left the finite volume at time $T$. It follows that, the lower
the energy, the larger $T$ will be required for the splitting to hold. Although each 
individual $\chi_\vk$ extends over the complete space, the scattering wave-packet
is localized outside $R_c$ up to a small error that quickly decays with growing $T$.
The scattering amplitudes $b_\vk$ can be obtained as
\beq
e^{iT\vk^2/2}b(\vk)=\l \psi_k | \Psi_{\rm s}(T)\r 
\approx\l \psi_k |\theta(R_c)| \Psi_{\rm s}(T)\r = \l \chi_k |\theta(R_c)| \Psi_{\rm s}(T)\r.
\eeq
Here we introduced the notation
\beq
\l \psi_k |\theta(R_c)| \Psi_{\rm s}(T)\r:=\int_{|\vr|>R_c}d\up3r\, \psi^*_\vk(\vr)\Psi_{\rm s}(\vr,T)
\eeq
The substitution of the scattering solution with a plane wave $\psi_\vk\to\chi_\vk$ 
in the last step uses the asymptotic behavior (\ref{eq:asymptotic}).
It is exact, when all interactions vanish beyond $R_c$.

For converting the above matrix element to a time-integral over surface values, we must assume that
we know the time-evolution of the particle {\em after it has passed through the surface}. For that we
assume that there is a ``channel Hamiltonian'' $H_c(t)$ such that
\beq
H_c(t)=H(t)\quad {\rm \,for\, } |\vr|>R_c\,{\rm \, and }\,\forall t. 
\eeq
In case of a short range potential $V(\vr)\equiv0$ for $|\vr|>R_c$ this is {\em exactly} fulfilled by the Hamiltonian for 
the free motion in the laser field
\beq
H_c(t) = \frac12 [-i\vnabla-\vA(t)]^2,
\eeq
 where $\vA(t)=-\int_{-\infty}^t \vEf(t') dt'$ for the dipole field $\vEf(t)$.
The Volkov solutions
\beq
\chi_\vk(\vr) = (2\pi)^{-3/2}e^{-i\Phi(\vk,t)}e^{i\vk\cdot\vr},\qquad \Phi(\vk,t)=\frac12\int_0^t d\tau [\vk-\vA(t)]^2
\eeq
solve the TDSE
\beq
i\ddt|\chi_\vk(t)\r=H_c(t)|\chi_\vk(t)\r.
\eeq
We can now write
\bea
\nonumber
\lefteqn{\l \chi_k(T) |\theta(R_c)| \Psi_{\rm s}(T)\r
=\int_0^T dt \ddt \l \chi_k(t) |\theta(R_c)| \Psi_{\rm s}(t)\r}
\\\nonumber\qquad\qquad&=i\int_0^T dt \l \chi_k(t) |H_c(t)\theta(R_c)-\theta(R_c)H(t)| \Psi_{\rm s}(t)\r
\\\label{eq:amplitude}&=i\int_0^T dt \l \chi_k(t) |\left[-\frac12\Delta+i\vA(t)\cdot\vnabla,\theta(R_c)\right]| \Psi_{\rm s}(t)\r
\eea
The commutator vanishes everywhere except on the surface $|\vr|=R_c$.
Assuming linear polarization in $z$-direction $\vA(t)=(0,0,A(t))$, it can be written 
in polar coordinates $(r,\theta,\phi)$ as 
\bea
\lefteqn{\left[-\frac12\Delta+iA(t)\partial_z,\theta(R_c)\right]=}\nonumber\\
&&\qquad-\frac12 r^{-2}\partial_r r^{2}\delta(r-R_c)-\frac12 \delta(r-R_c)\partial_r-iA(t)\cos\theta\delta(r-R_c)
\eea
With this, the volume integral over the space covered by the solution at time $T$ has been converted to a
time-integral up to $T$ and a surface integral over $|\vr|=R_c$.

We would like to remark that without time-dependence we can make one more step, as then the Volkov phase reduces to
$\Phi(\vk,t)=t\vk^2/2$ and time-integration turns into the time-energy Fourier-transform of the surface integral, 
which connects to the well-known results of, e.g.,  Ref.~\cite{tannor93:spectra}. 

\section{Finite element discretization and irECS}

For the efficient use of t-SURFF, a reliable mechanism
for truncating the solution outside the finite volume without generating reflections or other
artefacts is needed. Commonly complex absorbing potentials are employed for that purpose 
(for a recent review, see \cite{muga04:cap}).
We use infinite range exterior scaling (irECS) introduced and discussed in detail 
in reference \cite{scrinzi10:irecs}. Here we only briefly summarize the procedure. 

Exterior complex scaling (ECS) consists in making an analytic continuation of the Hamiltonian
by rotating the coordinates into the upper complex plane beyond the ``scaling radius'' $R_0$:
\begin{equation}
  \label{eq:ecs-definition}
  \vr\to \vr_{\Rth}=\begin{cases}
\vr{\rm \,for\, }|\vr|\leq R_0\\
\frac{\vr}{|\vr|}\left[R_0+e^{i\theta}(|\vr|-R_0)\right]{\rm \,for } |\vr| >R_0
\end{cases}
\end{equation}
The resulting complex scaled Hamiltonian $H_{\Rth}$ can be used in a complex scaled time-dependent
Schr\"odinger equation 
\beq\label{eq:tdse-scaled}
i\ddt\Psi_{\Rth}(t)=H_{\Rth}(t)\Psi_{\Rth}(t). 
\eeq
It was observed in \cite{scrinzi10:irecs} that for the velocity form of the TDSE, 
in the unscaled region   $|\vr|<R_0$ exact and  complex scaled solution agree:
$\Psi_{\Rth}(\vr,t)\equiv\Psi_v(\vr,t)$.
Mathematical proof is absent, but numerical agreement can be pushed to machine precision
with relative errors $\sim 10^{-14}$. 

These high accuracies can be reached with little effort by   
infinite range ECS (irECS), where very few $\sim 20$ discretization coefficients are needed at
radii $>R_0$. It consists in using an expansion into spherical harmonics and discretize the
radial parts by high order finite elements. As the last element, irECS uses 
the {\em infinite} interval $[R_0,\infty)$ and functions of the form $L_n(2\alpha r)\exp(-\alpha r)$ 
for its discretization, where $L_n$ are Laguerre polynomials. The idea is that low momentum / long de-Broglie wave length
electrons need rather large absorption ranges to become absorbed, but that the solution
at large distances is rather smooth. The more oscillatory short wave length content of the wave function 
becomes absorbed within a few oscillations.
The exponentially damped functions turned out be very efficient in emulating that behavior: in most
cases, about 20 functions at $|\vr|>R_0$ are sufficient for complete absorption.
Typical finite element orders used are 15 to 25. The comparatively high order further enhances 
efficiency of irECS.

Although in irECS there is no strict box boundary, the number of discretization
points and the phase space volume that can be represented remain finite. 
There is a correlation between position discretization and momentum discretization
in irECS: large electron momenta can only be represented
in the unscaled region and at the beginning of the scaled region. At long distances 
into the scaled region, only gentle oscillations and therefore low momenta can be represented
until the wave function dies out exponentially. 

As the finite element basis is local, the discretization coefficients are approximately associated
with positions in space. In fact, we can establish a one-to-one relation between the $M$ functions belonging
an element and the function
values at $M$ different points within that element. In that sense we will refer to the discretization coefficients as 
``discretization points'', emphasizing the locality of the finite element functions.

The exact choice of the scaling parameters $\theta, R_0$ and $\alpha$ is not critical. In Ref.~\cite{scrinzi10:irecs}
we found little variation of accuracy for values in the intervals  $\theta\sim[0.5,0.9]$ and $\alpha\sim[0.2,0.6]$.
Variations of the results with $R_0$, as a rule, reflect the spatial discretization error of the calculation. 
We found this behavior confirmed also for calculation of the photo-electron spectra presented here.
The plots shown below were all calculated with $\theta=\alpha=0.5$.

\section{Photo-electron momentum spectra  for a short range potential}

We solve the time-dependent Schr\"odinger equation in velocity gauge
\begin{equation}\label{eq:tdse-v}
i\ddt \Psi_v(\vr,t) = \left\{\frac12[-i\vnabla-\vA(t)]^2 + V(\vr)\right\}\Psi_v(\vr,t)
\end{equation}
where we first use the short-range ``Coulomb'' potential 
\beq\label{eq:short}
V(r)=\begin{cases}
c\left[-1/r-r^2/(2R^3)+3/(2R)\right]{\rm \,for\, } r\leq R\\
0  {\rm \,for\, }r>R.
\end{cases}
\eeq
With $R=20$ and an effective charge $c=1.1664$ the ground state energy is $-0.5$. 
The laser pulse is linearly polarized in $z$-direction with the vector potential
\beq
A_z(t) = \frac{\Ef_0}{\omega}\cos^2(\frac{\pi t}{2T}) \sin(t\omega).
\eeq
We choose parameters  $\omega=0.057$ and $\Ef_0=0.0755$ corresponding to laser wave length of 800 nm and 
peak intensity $2\times 10^{14}W/cm^2$. $T$ is the full width at half maximum (FWHM) of the
vector potential, total pulse duration is $2T$.

\begin{figure}
\begin{center}
\includegraphics[width=16cm]{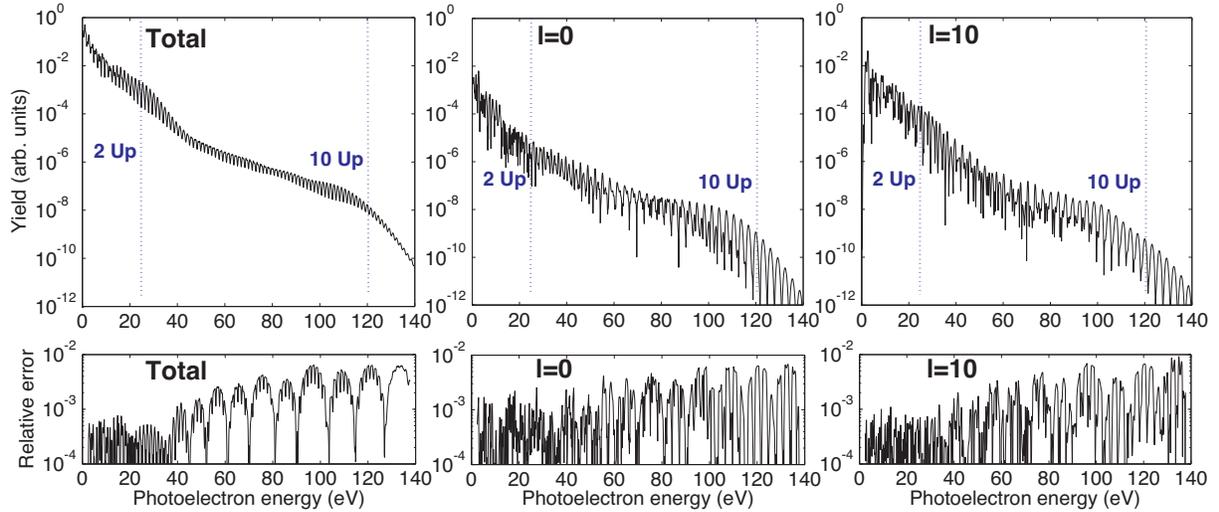}
\caption{\label{fig:short_20}
Upper panels: total and partial wave photo-electron spectra obtained 
for the smoothly truncated Coulomb potential Eq.~(\ref{eq:short}).
The lower panels show the relative errors for each spectrum  
according to Eq.~(\ref{eq:relative-error})  of a calculation with 
with only 90  discretization points per angular momentum comparing to a fully converged calculation.
Pulse parameters:  $\lambda=800\,nm$, $T=5$ optical cycles, intensity $= 2\times 10^{14}W/cm^2$.
}
\end{center}
\end{figure}

Figure~\ref{fig:short_20} shows the total and partial wave photo-electron spectra
for potential range $R=20$ and $T=5$ optical cycles. At these parameters,
more than 90\% of the electrons get detached. For accuracy $\lesssim 1\%$  
up to energies of $10\,U_p\approx 120\, eV$ we need $L_{\max}=30$ partial waves with only 90 
radial discretization points.
We define the error relative to an accurate reference spectrum $\sigma_{{\rm \,ref}}$ as
\begin{equation}
  \label{eq:relative-error}
  \mathcal{D}(E) = 
\frac{|\sigma(E) - \sigma_{{\rm \,ref}}(E)|}{\max(\sigma(E),\l\sigma(E)\r_{\delta E})},\quad
\l\sigma(E)\r_{\delta E}:=\frac{1}{2\delta E}\int_{E-\delta E}^{E+\delta E}dE'\sigma(E').
\end{equation}
Including in the denominator the average over the interval $[E-\delta E,E+\delta E]$
suppresses spurious spikes in the error due to near-zeros of the spectrum. We choose $\delta E = 0.05 a.u.\approx 1.5\,eV$,
which at 800 nm corresponds to averaging over about 2 photo-electron peaks.

In the unscaled region we use 60 points, 30 points are located in $|\vr|>R_0$. 
The accuracy estimate shown in Fig.~\ref{fig:short_20}
is obtained by comparing to a fully converged calculation. When we increase the number of points to 
180, the error drops to $\lesssim 10^{-3}$
The increase of relative errors with energy can be attributed to the decrease of the signal: 
note that from 0 to $10U_p$ the spectrum drops by more than 5 orders of magnitude.  

As a consistency check, we find that spectra up to $10\,U_p$ 
computed with largely different surface radii $R_c=21$ and 29 coincide 
within relative accuracies of better than $\lesssim 10^{-3}$. 
As beyond $R_c$ we assume the exact Volkov
propagation, this demonstrates that the solution of the TDSE and the integrals
(\ref{eq:amplitude}) are correct and that also reflections are suppressed
on at least that level of accuracy. 

The calculated spectra are independent of the complex scaling parameters:
over the ranges $R_0\in[20,30]$ and $\theta=[0.4,0.7]$ results vary 
by less than $10^{-3}$.
The lower limit for $R_0$ is not dictated by complex scaling: rather, as 
we want to obtain exact results, we must pick up the exact wave-function
outside the range of the potential $R_c\geq R=20$, and therefore also $R_0\geq R_c>R$. 
Already with these 
parameters the quiver amplitude, i.e. the excursion of free electrons  in the 
laser field $\Ef_0/\omega^2\approx 23$ reaches beyond $R_0$ and into the 
complex scaled region. This confirms an earlier observation that 
the dynamics is correctly reproduced also in the complex scaled region 
\cite{scrinzi10:irecs}.

\begin{figure}
\begin{center}
\includegraphics[width=14cm]{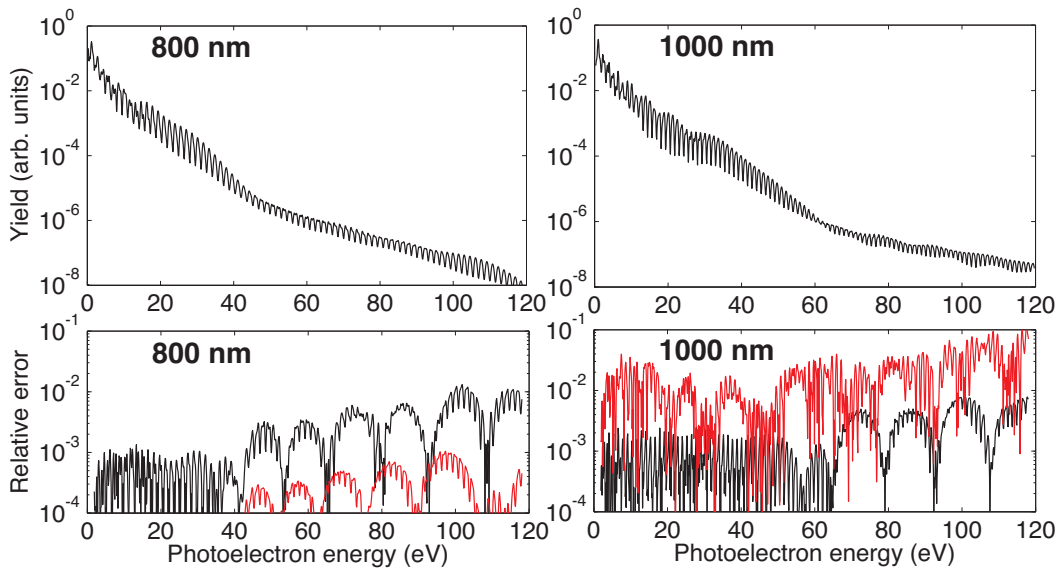}
\caption{\label{fig:quiver}
Required box sizes depend on the quiver amplitude. Left panels: spectrum (top) 
for a 5 cycle FWHM pulse at 800 nm and $\Ef_0=0.0755$ (intensity $2\times 10^{14}W/cm^2$).
Accuracies (bottom) are calculated with 
75 (black line) and 100 (red line) discretization points. Errors $\lesssim1\%$ relative to an accurate reference calculation
are reached with 75 points. At 100 points the error is $\lesssim 10^{-3}$. Right panels:
spectrum (top) and accuracies (bottom) for 5 cycles FWHM at 1000~nm and the same intensity. 
Because of the larger quiver amplitude, a $\sim 60\%$ larger box with 120 points is needed for $<1\%$ accurate results. 
When using a smaller box with 120 points at a 30\% higher density 
errors increase to near 10\% (red line).
}
\end{center}
\end{figure}

For correct electron spectra, the effective box
size of the combined unscaled and scaled regions must be large enough
to accommodate the quiver motion. To study this further, we use a 
somewhat shorter effective range of $R=15$ and choose $R_0=R_c=R$.
From Fig.~\ref{fig:quiver} we see that with only 45 discretization point in the 
unscaled region $r<R_0$ 1\% accuracy is reached at the same laser parameters 
as before. With 30 points for absorptions we have a total of 75 points.
Note that the quiver radius of $\approx 23\, a.u.$  
now reaches rather deep into scaled region but still fits into 
the total box. 

Keeping the intensity fixed, a longer wave length of 1000~nm leads to 
a larger quiver radius of $\approx36\,a.u.$. 
We expect to need a factor $(1000/800)^2\approx 1.6$
more discretization points. 
Indeed, we can reach $\lesssim1\%$ accuracy with
total of 120 points in this case (Fig.~\ref{fig:quiver}). 
All additional coefficients and at least half of the quiver motion 
now are located in the scaled region. 
Note that also $U_p$ and with it the peak momentum grows with wave length: 
for describing the energies $>120\, eV$
we would also need to increase the density of points by 25 \%.
However, just increasing the number of points without increasing the box size
gives incorrect results: it appears that we need to accommodate the full
quiver motion up to $\sim 36\,a.u.$ in the simulation box. 
  
\section{Photo-electron momentum spectra for  the hydrogen atom}

As always in scattering problems, the long-range nature of the Coulomb potential 
introduces extra mathematical and practical complications. Considering t-SURFF,
there is no surface radius $R_c$ such that the Volkov solutions become exact.
In addition, the Rydberg bound states extend to arbitrarily large distances. 
We will discuss below how these problems can be eliminated 
with moderate extra computational effort.
Here we present the pragmatic solution of  using 
larger $R_c$ such that the remaining error due to the presence of the Coulomb 
potential becomes acceptable. 

\begin{figure}
\begin{center}
\includegraphics[width=12cm]{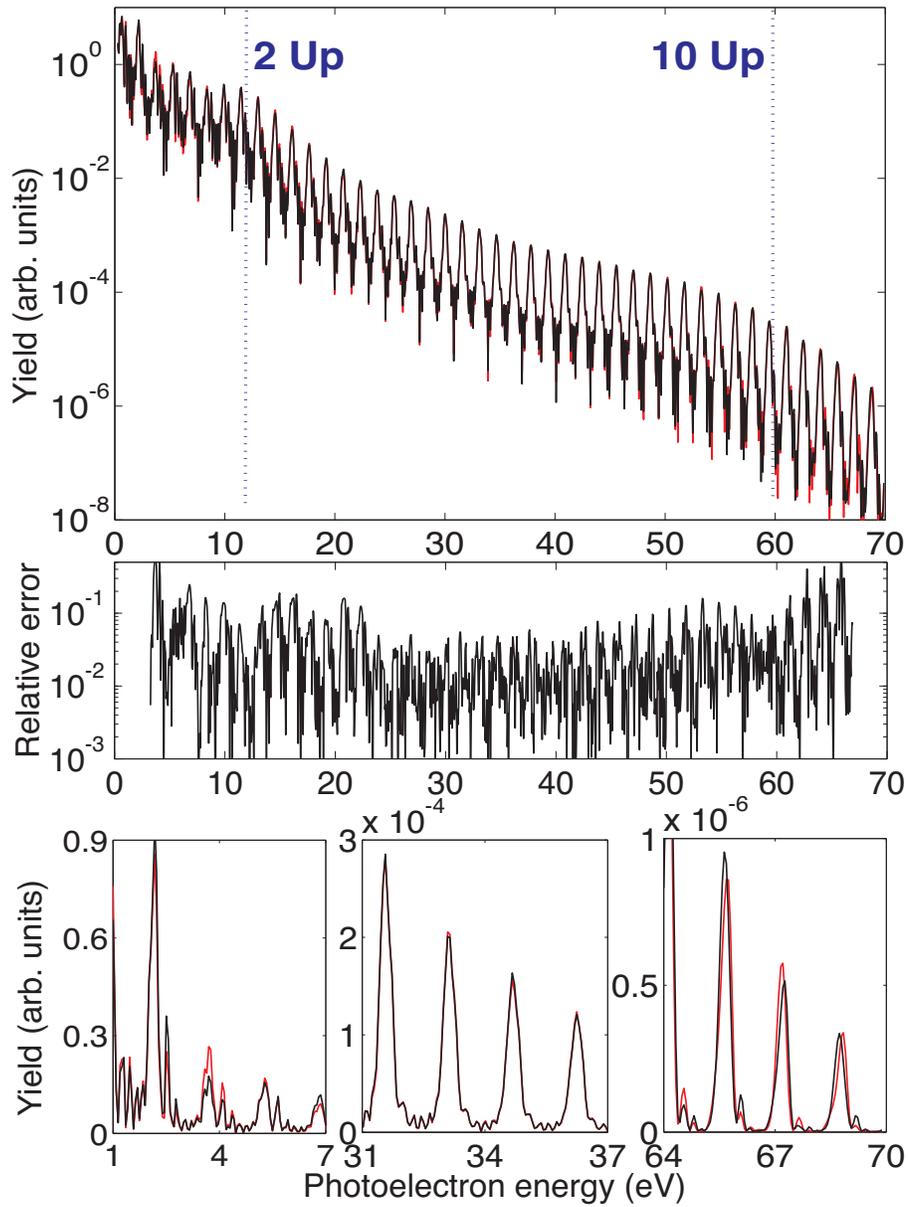}
\caption{\label{fig:coulomb}
Photo-electron energy spectra for the hydrogen atom (upper panel) obtained with surface radius $R_c=110$ (black) 
and $R_c=140$ (red). Middle panel: error estimate by comparing the spectra
according to Eq.~(\ref{eq:relative-error}). Lower panels: blow-up of the spectra from the upper panel on a linear scale.
Pulse parameters:  $\lambda=800\,nm$, $T=20$ optical cycles, intensity $= 10^{14}W/cm^2$.
}
\end{center}
\end{figure}
Fig.~\ref{fig:coulomb} shows a spectrum calculated 
for the Hydrogen atom with a FWHM $T=20$ optical cycle pulse at 800 nm wave length
and peak intensity $10^{14}W/cm^2$. All discretization errors can be controlled in 
the same way as discussed for the short range potential. The error is dominated
by the dependence on the surface radius $R_c$: on an absolute (logarithmic)
scale, two calculations with $R_c=110$ and $R_c=140$ are hardly discernable. 
The error level of the calculation with $R_c=110$ and 180 discretization points 
is $\lesssim 10\%$ and decreases slowly as $R_c$ increases. 
In the linear plot of the spectra (lowest panels of Fig.~\ref{fig:coulomb}) we see
that the largest errors occur at lower energies due to the larger influence of the weak 
Coulomb tail on low energy scattering states. Agreement at intermediate energies
is near perfect. The increase of relative errors at the highest 
energies is due to a displacement of peaks caused slightly incorrect dispersion
due to spatial discretization.

\begin{figure}
\begin{center}
\includegraphics[width=16cm]{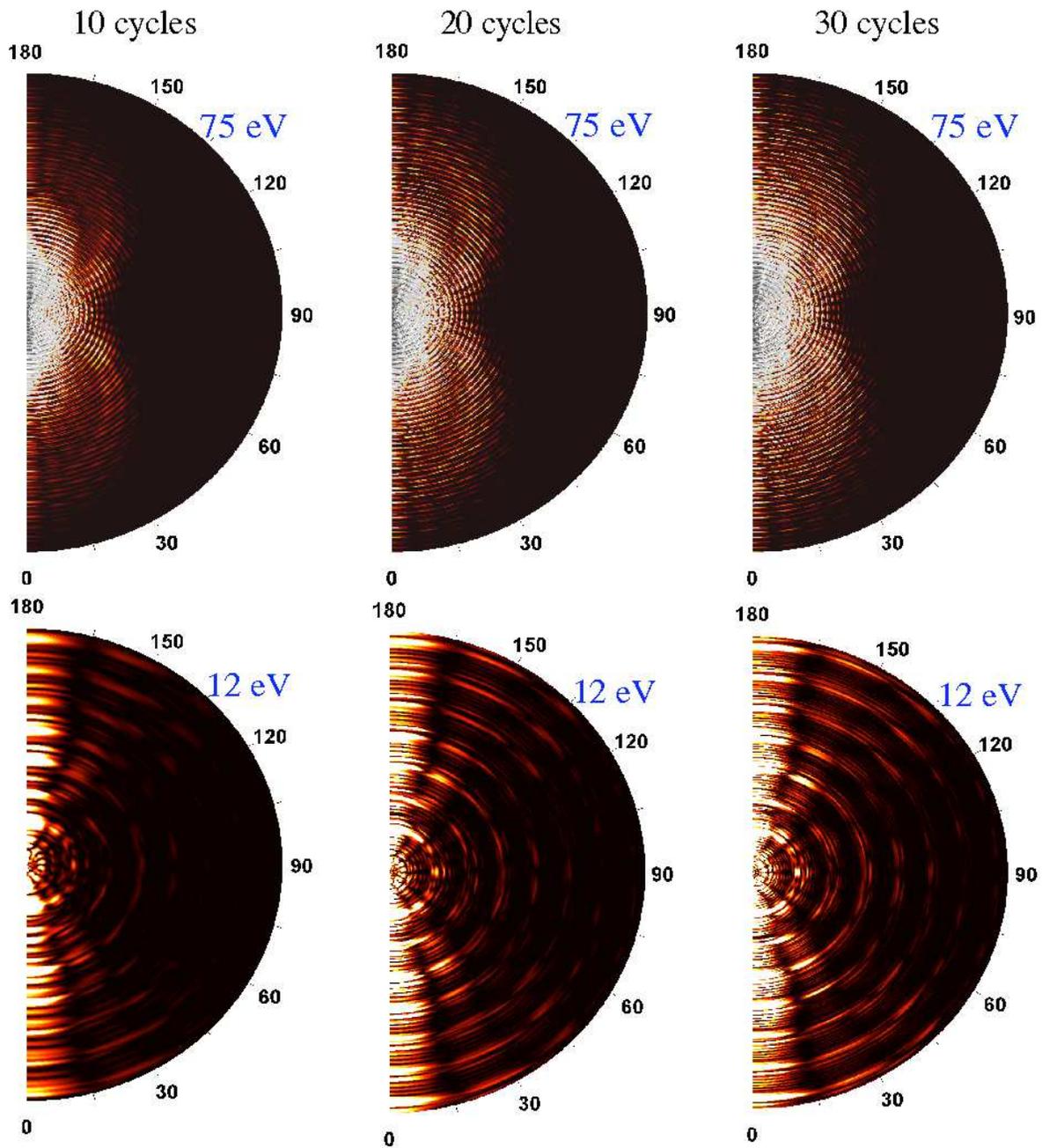}
\caption{\label{fig:2d}
Angle resolved photo-electron energy spectra for the hydrogen atom at $\lambda=800\,nm$, intensity $10^{14}W/cm^2$ and FWHM pulse durations $T=10$, 20, and 30
(upper panels). In the lower panels, the region up to $2U_p\approx 12\,eV$ is enlarged.
}
\end{center}
\end{figure}
Fig.~\ref{fig:2d} shows angle-resolved photo-electon sprectra for FWHM durations of T=10, 20 and 30 optical cycles. 
In the region up to $10\,U_p\approx 60 \,eV$ circular structures with their centers offset along the polarization
axis are clearly distinguishable, best visible at
the shortes pulse $T=10$. These structures were first explained in Ref.~\cite{morishita08:rescattering} and are due to
re-scattering. The intensity around each circle
is related to the electron-ion scattering differential cross section \cite{chen09:rescattering}. 
In the enlarged plots of the energy region up to energies of 
$2\,U_p\approx 12 \,eV$ we reproduce rings and fan-like structures that are related to particular partial waves, 
as has been extensively discussed in Ref.~\cite{morishita07:rescattering}.
Sub-structures between the photo-electron energy peaks are clearly resolved. These
are related to the pulse envelope: the spacing decreases and number of peaks increase as the pulse-duration 
increases. In different wording one can say that they are caused by interference between rising and 
trailing edges of the pulse \cite{telnov95:photoelectrons}.

\begin{figure}
\begin{center}
\includegraphics[width=12cm]{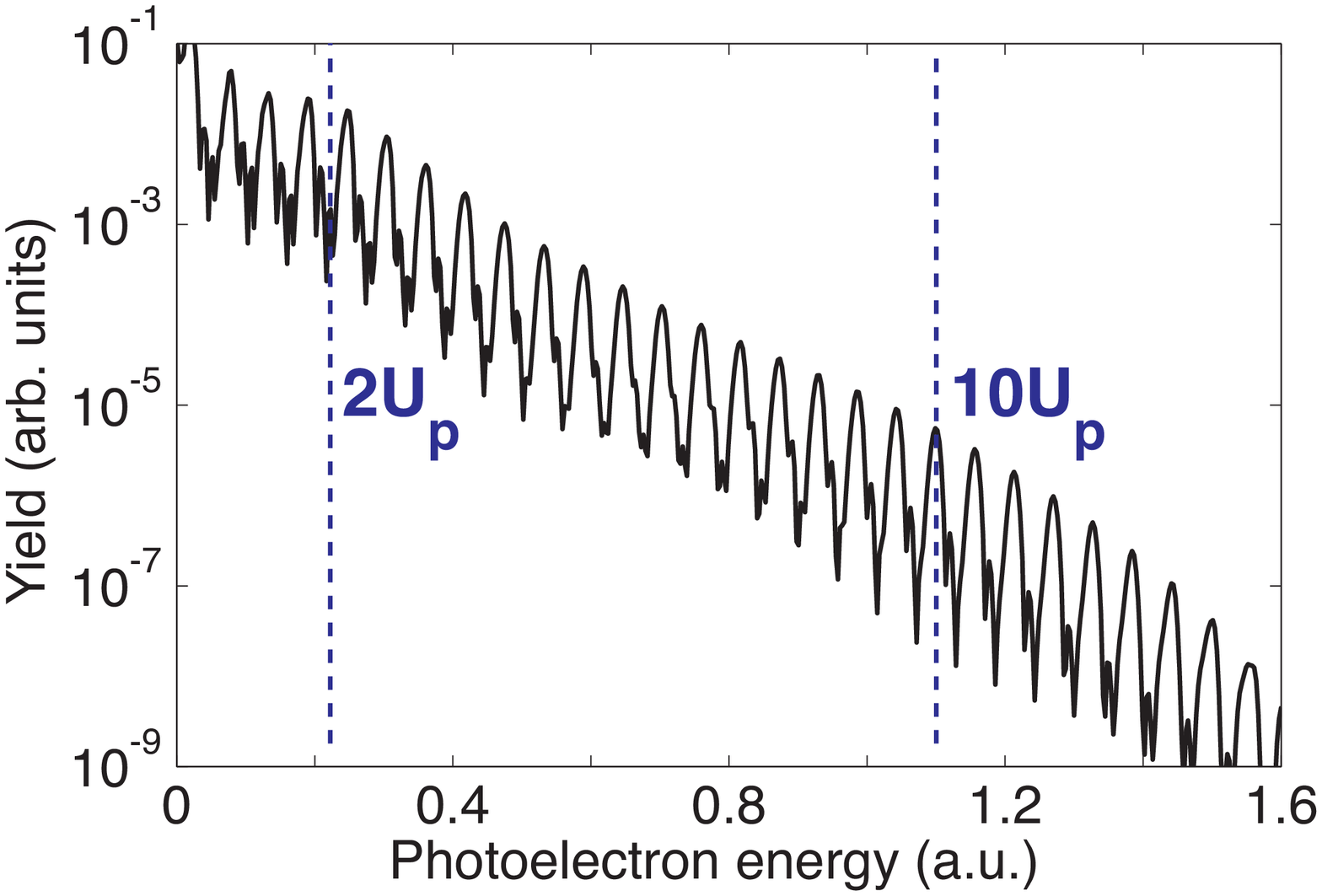}
\caption{\label{fig:zhou}
Photo-electron energy spectrum for a FWHM $T=10$ pulse at 800 nm and peak intensity $I=5\times 10^{13}W/cm^2$.
Results obtained with 192 radial discretization points are accurate to a few percent and 
never exceed $10\%$ in the whole range shown.  
The pulse parameters agree with those used for Fig.~2 in Ref.~\cite{zhou11:p-space}.
Near $10\,U_p$ there apears a striking qualitative difference between our result and Ref.~\cite{zhou11:p-space}.
(See text for a discussion.)
}
\end{center}
\end{figure}
For comparison with results reported in \cite{zhou11:p-space}, we also computed the spectrum at the lower intensity 
of $5\times 10^{13}W/cm^2$ and $T=10$ cycles FWHM (corresponding 20 cycles total pulse duration).
Fig.~\ref{fig:zhou} shows our result for the total photo-ionization spectrum obtained with only 192
discretization points and 25 angular momenta. 
We have estimated the accuracy of our results to be $\lesssim10\%$ throughout the spectrum
using the same procedures as for Fig.~\ref{fig:coulomb}. 
Our result qualitatively differs from Fig.~2 in Ref.~\cite{zhou11:p-space}, where
a surprsing irregularity appears near $10\,U_p$, while no such structure exists in our sprectrum.
Unfortunately no detailed discussion of accuracy or convergence is given in \cite{zhou11:p-space}.
One possible source of the discrepancy is insufficient discretization.
With peak energy of $40\,U_p\sim 4 a.u.$ included in the calculation the maximal momenta are 
$p_{\max}\sim 2.8\,a.u.$. During 20 optical cycles (2200 a.u.\ of time), these electrons move to distances
$\sim 6000\,a.u.$. With the 2000 discretization points used in \cite{zhou11:p-space}
one gets average momentum grid spacing of $\Delta p=p_{\max}/2000$ and an effective 
spatial box size $\sim2\pi/\Delta p \approx 4500$, which appears somewhat below the necessary 
limit. However, the uneven distribution of grid points and additional spectral cuts in the 
energy domain used in \cite{zhou11:p-space} make it difficult to carry this analysis further.
Rather, a systematic convergence study would be required.   
The present number of discretization points also compares well to the $1000\sim2000$
discretization coefficients used in \cite{hamido11:time-scaling} at photon energies $0.3\sim0.7\,a.u.$.
The results of the benchmark calculations Refs.~\cite{cormier97:spectra} obtained with box size of 
3000 atomic units and about 4000 discretization points at somewhat shorter wave length of 620~nm 
could be reproduced using only 200 radial discretization point up to 
energies of $10\,U_p$, except at the lowest energies, where our method is limited by use of Volkov solutions
beyond the surface radius $R_c\lesssim 150 a.u.$.

\section{Extensions of the method}

\subsection{Handling long-range potentials}

In the section above we were able to obtain good results for quite demanding laser parameters using 
only $R_c\approx 140$. Still, the problem remains big and in particular
when considering extension of the approach to multi-electron systems, multi-dimensional simulation volumes
would quickly exhaust computational resources. In turn, a reduction of $R_c$ to the necessary
minimum of $20\sim 50$ set by the quiver radius, would constitute an essential gain, possibly deciding 
about the feasibility of the multi-electron calculation. Here we show how to correctly handle
long-range potentials in t-SURFF.

For computing exact spectra, we must know the exact solution of the TDSE beyond $R_c$.
In the derivation of Eq.~(\ref{eq:amplitude}) we have only used $H(t)\equiv H_c(t)$ 
for $|\vr|>R_c$ and that we can by some means obtain accurate solutions $\chi_\vk(\vr,t)$ for the TDSE 
with $H_c(t)$. A suitable $H_c$ is obtained by using in Eq.~(\ref{eq:tdse-v}) the potential
\beq
V_c(r)=\begin{cases}
-1/R_c{\rm \,for\, } r\leq R_c\\
-1/r  {\rm \,for\, } r>R_c.
\end{cases}
\eeq
Partial wave solutions $\phi_{k,l}(r)$ for the field-free case $\vA\equiv0$ with $V_c$ are 
the spherical Bessel functions up to $R_c$ which are connected smoothly to values and 
derivatives of regular and irregular Coulomb functions in the region $r>R_c$, where one
must pay attention to proper $\delta$-normalization. With non-zero field $\vEf(t)$, no
exact solution is knwon. We found that a simple Coulomb-Volkov
approximation \cite{duchateau02:coulomb-volkov} for the time-dependence is insufficient: 
we could not observe any acceleration of convergence by replacing the plane waves of the Volkov solution
with the field-free scattering solutions.
Lacking a reliable approximation for the scattering solution, we must solve the TDSE 
for the $\chi_\vk(\vr,t)$ with the final condition
\beq
\chi_{k,l}(r,T)=\phi_{k,l}(r).
\eeq
As the potential is weak, little actual scattering occurs and accurate solutions can be
obtained by expanding into  $\phi_{k',l'}(r)$ for small intervals around the asymptotic radial and 
angular momenta,  
$k'\in[k-\Delta k,k+\Delta k]$ and $l'\in[l-\Delta l,l+\Delta l]$. 
Numerical results using this procedure will be presented elsewhere.

Finally there remains the poblem that Rydberg states may become occupied, which have non-neglible
amplitude at $r=R_c$. The effect of bound states on the integrals (\ref{eq:amplitude}) is
slow, oscillatory convergence to the asymptotic value at $T\to\infty$. Again, there is an efficient and rather
pragmatic solution to this problem: by averaging the value over several optical cycles, we find that
convergence is speeded up and the reported accuracies are reached quickly. 

If the Rydberg states are known exactly, we can remove them after the pulse is over.
This removes all oscillations and the asymptotic value is reached rapidly.
For the procedure we define the projector onto the (field free) Rydberg states $|n\r$
and its complement as 
\beq
P:=\sum_n |n\r\l n|{\rm \, and \,} Q:=1-P.
\eeq
A simple calculation shows that the spectral amplitude with the Rydberg states removed is
\bea
\lefteqn{\langle \chi_\vk( T) | \theta(R_c) Q | \Psi(T) \rangle=
i\int_{-\infty}^{T_0}dt \langle \chi_\vk(t)| [H_c(t),\theta(R_c)] |\Psi(t) \rangle+}\nonumber\\
&\qquad& i\int_{T_0}^{T}dt \langle \chi_{\vk}| [H_c(t),\theta(R_c)]Q |\Psi(t) \rangle
-\langle \chi_\vk(T_0)|\theta(R_c) P|\Psi(T_0) \rangle ,
\label{eq:project}
\eea
where $T_0$ is any time after the end of the pulse.
If high precision is wanted and if long time-propagation is costly, the extra effort of implementing
the explicit projection (\ref{eq:project}) may be justified.

\subsection{Single-ionization of multi-electron systems}

The procedure can be extended to describe single-ionization of multi-electron
systems. The main new feature here is that we have several ionization channels, depending
on the state in which the ion is left behind. The spectral density in a specific channel $c$ 
can be written, as before, as
\beq
\sigma_c(\vk)=|\l \chi_{c,\vk}(T) | \theta(R_c) |\Psi_{\rm s}(T)\r|^2.
\eeq 
The asymptotic channel wave function $ \chi_{c,\vk}(t)$ fullfills the channel TDSE
\beq\label{eq:channel}
i\ddt \chi_{c,\vk}(t) = H_c(t)  \chi_{c,\vk}(t),
\eeq
with the channel Hamiltonian 
\beq
H_c(t) = \frac12 [-i\vnabla-\vA(t)]^2\otimes H_{ion}(t)
\eeq
where for simplicity we neglect the ionic Coulomb potential.
Note that we do not need to explicitly anti-symmetrize $\chi_c$, if $\Psi_{\rm s}$ is anti-symmetric.
A general solution of (\ref{eq:channel}) has the form
\beq
 \chi_{c,\vk}(t)= (2\pi)^{-3/2}e^{-i\Phi(t)}e^{i\vk\vr}\otimes \phi_c(t) ,
\eeq
where $\phi_c(t)$ solves the ionic TDSE with Hamiltonian $H_{ion}(t)$ and 
a final state condition that specifies the ionic state $c$ of the channel:
\beq
H_{ion}(T) \phi_c(T) = E_c \phi_c(T).
\eeq
Assuming that double-ionization is negligible, the further steps for computing the spectral amplitude
as a time-integral are the same as in the single-electron case. In addition to 
$\Psi_{\rm s}(t)$ we must also compute $\phi_c(t)$. Values and derivatives of
the $c$-channel scattering wave function $\varphi_c(\vr,t):=\l \phi_c(t) | \Psi_{\rm s}(t)\r$ at
$|\vr|=R_c$ can be stored for a sufficiently dense time-grid. 
Because of the stronger binding of electrons in the ion, computing the ionic wave function 
$\phi_c(t)$ usually requires much xsless effort than obtaing $\Psi_{\rm s}(t)$.  

\subsection{Double-ionization}
For double-ionzation spectra, we must know the two-electron solution in the
asymptotic region. This may be less difficult than what it appears at first glance.
Neglecting  the ionic potentials, we have a Volkov solution for the center of mass coordinate and
Coulomb waves for the relative coordinate. What is left to do is to match 
that solution with an accurate solution on a five-dimensional surface, 
where all reflections from the simulation box boundaries 
are carefully suppressed. While this procedure has not been worked out in detail, it may 
be well feasible and bring IR two-electron spectra from realm of herioc super-large scale calculations
\cite{taylor03:780nm} to managable size, allowing systematic studies.

\section{Conclusions and outlook}

We have shown that the unfavorable scaling for the computation of photo-electron
spectra with laser wave length can be largely overcome with t-SURFF, which picks up
the exact solution at some finite surface and beyond that surface exploits 
knowledge about the long-range behavior of the solutions of the TDSE. The reduction 
of problem size is particularly striking for atomic binding potentials with finite range, when the Volkov-solutions 
become exact at distances where the potential is zero. In that case, the box sizes can be reduced to 
approximately the range of the potential plus the electron quiver amplitude. Electrons that
move beyond that range will never scatter and will exactly follow the Volkov solution.
For our parameters, the wave function expands to several thousand atomic units during the pulse
and correspondingly large boxes would be needed, if a spectral analysis of the wave function
were performed after the end of the pulse. In contrast, we could present $\lesssim1\%$ accurate spectra up to 
energies of $120\,eV$ using a box size of only about 30 atomic units and as little as 75 discretization 
points per partial wave. With only a few more points, much higher accurcacies can be reached. 

Instrumental for the application is the traceless absorption of the wave function beyond the surface,
which is provided by the irECS method introduced in a preceding publication \cite{scrinzi10:irecs}. 
The good performance of irECS for one-dimensional wave functions and 
for high-harmonic signals from three-dimensional calculations presented in \cite{scrinzi10:irecs}
could be confirmed also for the much more delicate observable of angle-resolved photo-electron spectra.

For the long-range Coulomb potential, the Volkov solutions are not exact asymptotically,
let alone at any finite distance. In physical language, the electron will scatter in the 
long-range tail of the Coulomb potential and pick up more energy from the laser field and
therefore we cannot predict its final energy before the
pulse is over. An attempt to approximate the asymptotic behavior by Coulomb-Volkov instead 
of the pure Volkov solutions was futile. 
In a pragmatic approach, we could show that with a surface at distances of
$R_c=100\sim 140$ atomic units, still impressively accurate spectra can be obtained using Volkov
solutions as approximation to the exact asymptotic solutions. 

For single-electron systems and with moderate
accuracy requirements, simulation volumes on the scale of $\sim 100$ atomic units are quite acceptable. 
For experimentally interesting multi-electron systems, we have proposed to further reduce box-sizes
by  solving the asymptotic laser-assisted scattering problem numerically. 
This may be done efficiently for an asymptotic Hamiltonian that only includes
scattering at distances $>R_c$ and dismisses the main part of scattering from near the Coulomb singularity.
The need to solve this weak scattering problem enhances the complexity of coding and significantly 
increases computation times. However, for few-electron systems, this extra effort is far compensated
by an expected reduction to box sizes to as little as $20\sim50$ atomic units.
We have also formulated the extension of t-SURFF to single- and double-ionization of multi-electron systems.
A numerical demonstration of these methods will be the subject of future work.

In summary, t-SURFF, while producing highly accurate results, 
can reduce the box size for computing IR photo-electron spectra for single electron systems
by one order of magnitude or more.
This drastic reduction of box-sizes is particularly important for very long laser wave length 
and for multi-electron systems. For systems with two and more electrons, we believe, 
it opens a root to computing accurate IR  photo-electron momentum spectra.

\ack
This work has been supported by the Austrian Science Fund within the 
framework of the Special Research Program F41 Vienna Computational 
Materials Laboratory (ViCoM).

\section*{References}

\end{document}